\documentclass[]{mn2e}
\usepackage{graphicx}
\usepackage{amssymb,amsmath}

\newcommand{\sgr}{SGR~0418+5729}

\title[A    Magnetar    Strength    Surface   Magnetic    Field    for
SGR~0418$+$5729]{A  Magnetar Strength Surface  Magnetic Field  for the
  Slowly Spinning Down SGR~0418$+$5729}

\author[Tolga G\"uver,  Ersin G\"o\u{g}\"u\c{s}, \&  Feryal \"Ozel]{Tolga
  G\"uver$^{1}$\thanks{E-mail:tguver@email.arizona.edu},          Ersin
  G\"o\u{g}\"u\c{s}$^{2}$, \&  Feryal \"Ozel$^{1}$ \\ $^{1}$Department
  of Astronomy, University of Arizona, 933 N.  Cherry Ave., Tucson, AZ
  85721\\   $^{2}$Sabanc\i~University,  Faculty  of   Engineering  and
  Natural Sciences, Orhanl\i $-$ Tuzla, \.Istanbul, 34956 Turkey}

\begin{document}

\pagerange{\pageref{firstpage}--\pageref{lastpage}} \pubyear{2011}

\maketitle

\label{firstpage}

\begin{abstract}
  The  observed upper bound  on the  spin down  rate of  the otherwise
  typical   Soft   Gamma-ray  Repeater   \sgr\   has  challenged   the
  interpretation  of this source  as a  neutron star  with ultrastrong
  magnetic  fields.   Current limits  imply  a  dipole magnetic  field
  strength of  less than  $7.5 \times 10^{12}$~G  (Rea et  al.  2010),
  which  is  significantly  smaller   than  that  of  a  typical  SGR.
  Independent of the properties  inferred from X-ray timing, the X-ray
  spectra of neutron stars allow a measurement of their magnetic field
  strengths because  they are distorted  from pure blackbodies  due to
  the  presence  of  a  magnetic  field  in  a  radiative  equilibrium
  atmosphere.  In this paper, we model high signal-to-noise XMM-Newton
  spectra  of  \sgr\ to  place  constraints  on  the strength  of  the
  magnetic field  at the  surface of the  neutron star.   Our analysis
  shows  that neutron  star atmosphere  models with  moderate magnetic
  field  strengths  ($10^{12-13}$~G)  cannot  fit the  X-ray  spectra,
  whereas, models  with stronger magnetic  fields are able  to account
  for the  observations.  We  find that the  strength of  the magnetic
  field at the surface is $1.0 \times 10^{14}$~G. This value, although
  lower than  all of the  other SGRs analyzed  to date, is  still high
  enough to  generate the observed  X-ray bursts from the  source.  In
  connection to  the spindown limits, it also  implies a significantly
  non-dipolar structure of the  magnetic field. We discuss the results
  of our spectral modeling and compare them with other SGRs.
\end{abstract}


\section{Introduction}

Soft Gamma-ray Repeaters (SGRs)  are neutron stars that are identified
by  the   repeated  bursts   they  emit  in   hard  X-rays   and  soft
gamma-rays. During their burst  active phases, SGRs emit anywhere from
a few to thousands of short  bursts, typically lasting a fraction of a
second.   Energy released  during such  a  short time  is very  large,
ranging from $\sim 10^{37}$~erg  to 10$^{40}$~erg.  On rare occasions,
SGRs emit extremely energetic giant flares, lasting for a few hundreds
of seconds  and releasing  a total energy  in excess  of $10^{44}$~erg
(Palmer et al. 2005).

SGRs also  display persistent bright X-ray  emission with luminosities
of  the  order  of  L$_{\rm  X}\lesssim  10^{35}$~erg~s$^{-1}$,  which
significantly  exceeds the  spindown power  that can  be  generated by
these  slowly  rotating  neutron  stars.  Both  the  persistent  X-ray
emission and the energetic bursts led to the interpretation of SGRs as
extremely magnetized  neutron stars, or magnetars  (Duncan \& Thompson
1992; see Woods  \& Thompson 2006 for a  detailed review).  Within the
magnetar  paradigm,  the  decay   of  a  very  strong  magnetic  field
($10^{14}-10^{15}$~G)  powers   the  persistent  emission   from  SGRs
(Thompson \&  Duncan 1996), while  the observed bursts  are attributed
either to cracking  of the neutron crust that  is strained by magnetic
stress (Thompson \& Duncan 1995) or to magnetic reconnection (Lyutikov
2003).

Large   period  derivatives,   of   the  order   of  $\dot{P}   \simeq
10^{-11}$~s~s$^{-1}$ measured from numerous SGRs in the past indicated
large inferred dipole spindown fields $B_{\rm dip} = 10^{14} (P/5 {\rm
  s})^{1/2}  (\dot{P}/10^{-11}\,{\rm s}\,{\rm  s}^{-1})^{1/2}$~G, thus
lending  further support to  the magnetar  character of  these sources
(e.g., Kouveliotou  1998). When models of high  magnetic field neutron
star atmospheres and magnetospheres were  used to fit to the continuum
X-ray spectra of magnetars, these analyses also yielded magnetic field
strengths that  are comparable  to the inferred  dipole fields  with a
typical range of  2 $ -$ 5 $\times$ 10$^{14}$  Gauss (e.g., G\"uver et
al.\ 2007,  2008; \"Ozel, G\"uver,  \& G\"o\u{g}\"u\c{s}\ 2008;  Ng et
al.\ 2010; G\"o\u{g}\"u\c{s} et al.\ 2011).

The ubiquitous  presence of  large period derivatives  in SGRs  is now
being challenged by the  unusually small period derivative measured in
\sgr.  Early attempts to  obtain its period derivative, and therefore,
to  establish  its  inferred  dipolar magnetic  field  strength,  were
inconclusive (Woods et al.  2009;  Kuiper \& Hermsen 2009; Esposito et
al. 2010).  Using multi-satellite  observations spanning over 440 days
following  the  onset  of  the  outburst  from  this  source,  Rea  et
al. (2010)  recently reported  a 2$\sigma$ upper  limit to  the period
derivative, implying that the inferred dipolar magnetic field strength
of \sgr\ should be less than $7.5 \times 10^{12}$~G.

SGR 0418+5729  is a “regular” SGR  in every other way  and a transient
magnetar candidate (Rea  \& Esposito 2011): It was  discovered on 2009
June  5  by emitting  two  bursts detected  with  GBM  on board  Fermi
Gamma-ray Space  Telescope (van der  Horst et al.  2010).   The energy
released  by  these  two   events  were  modest,  totaling  $8  \times
10^{36}$~erg  and $4  \times 10^{37}$~erg  in the  20$-$200  keV band.
Rossi X-ray  Timing Explorer observations of the  source following the
discovery revealed  the 9.07~s  spin period (G\"o\u{g}\"u\c{s}  et al.
2009).   Detectors on  board Swift,  Fermi,  and RXTE  have allowed  a
growing number of  transient magnetars to be discovered  in the recent
years, which are first  detected via bursting activity and accompanied
persistent  flux   increase  up   to  several  orders   of  magnitude.
Persistent flux of these transient sources decreases back to quiescent
level ($\approx$ 10$^{-13}$ erg~cm$^{-2}$~s$^{-1}$) on a time scale of
months to  years (Rea  \& Esposito 2011).   X-ray output  variation of
\sgr\ also very much  resembles that of transient magnetar candidates:
the 1$-$10~keV flux rapidly increased in conjunction with bursting and
subsequently decayed by  a factor of 10 over a  time span of $\sim$150
days (Esposito et  al.  2010).  \sgr\ is located  towards the Galactic
anti-center and  likely to be in  the Perseus arm or  the outer-arm of
our Galaxy with a distance of  $\approx$2 kpc or higher (van der Horst
et al.\ 2010).

In  this paper,  we aim  to place  constraints on  the  magnetic field
strength of \sgr\  using X-ray spectroscopy.  To this  end, we analyze
the  high  signal-to-noise  X-ray  spectrum  of  \sgr\  obtained  from
XMM-Newton observations.  We use  a number of different X-ray spectral
models to constrain  the physical properties of the  neutron star.  In
particular,  we   employ  (i)  the   phenomenological  blackbody  plus
power-law model that  has been traditionally used to  fit SGR spectra,
(ii) the low-to-moderate magnetic field Neutron Star Atmosphere model,
or NSA  (Pavlov et al.  1995;  Zavlin, Pavlin, \&  Shibanov 1996), and
(iii)   the  high   magnetic  field   Surface  Thermal   Emission  and
Magnetospheric  Scattering  (STEMS)  model  (G\"uver,  et  al.   2007,
2008). We find that the thermal spectrum of \sgr\ is best described by
high  magnetic  field  model.   We  also compare  the  magnetic  field
determined  spectroscopically to  the dipole  magnetic  field inferred
from spindown.  Because  the spectroscopically determined field probes
the strength measured on the  neutron star surface, whereas the dipole
field is the  one inferred at the light  cylinder, this comparison can
be revealing for the field geometry of \sgr.

We introduce the  XMM-Newton data and our data  reduction procedure in
the  next section.   In  \S 3,  we  fit the  spectrum  with the  three
different continuum  models. Finally,  we discuss the  implications of
the inferred surface magnetic  field strength and compare our findings
for this source with those of  SGRs in general to address the possible
differences between them in the final section.

\section{Observations and Data Analysis}

In the period following its discovery through its X-ray/soft-gamma-ray
bursts,  \sgr\  could  not   be  observed  immediately  with  pointing
telescopes due to its  unfavorable sky location.  In particular, Swift
XRT observations  of the  source could start  about a month  after the
discovery, while  the XMM-Newton  observation was performed  about two
months  after.  Because  the X-ray  intensity of  the  source steadily
declined  following  the  outburst   (see  Figure  1  of  Esposito  et
al.~2010), only  the large collecting  area of XMM-Newton was  able to
provide sufficiently  high quality X-ray spectrum  for the measurement
of spectral parameters.  We, therefore, use in this study the archival
XMM-Newton  observation that  took place  on 2009  August 12  (Obs ID.
0610000601).  We  note that \sgr\  was still not in  quiescence during
the  XMM-Newton  observation  (Esposito  et  al.~2010).   During  this
observation, the  source count-rate as  observed with the  EPIC-pn and
MOS detectors  were, 1.35  and 0.5 c~s$^{-1}$,  respectively.  EPIC-pn
and MOS detectors were operated in small and partial window modes with
time  resolutions of  6 and  300 ms,  respectively, to  prevent photon
pile-up.

Part of the XMM-Newton observation, especially towards the end, was
severely affected by large particle flares. We are, therefore, able to
use $\approx$36~ks out of the total exposure time of 65~ks.  We
utilize data collected with both the EPIC-pn and EPIC-MOS cameras.

We used  the {\it epproc} and  {\it emproc} tasks for  the EPIC-pn and
EPIC-MOS data with the  Science Analysis Software (SAS) version 10.0.0
and the  latest available  calibration files as  of December  2010. We
extracted X-ray  spectra using a circular  region with a  radius of 32
arc-seconds  centered on  the source.   Similarly,  background regions
were selected  from a source free  region with a typical  radius of 50
arc-second.  We used the {\it rmfgen} and {\it arfgen} tools to create
response and  ancillary response files. We rebinned  the X-ray spectra
to have at least 30 counts per bin and not to oversample the intrinsic
energy resolution of the detectors by more than a factor of 3.

\section{X-ray spectral analysis and results}
\label{spec}

We fit the data with XSPEC version 12.5.1n (Arnaud, 1996).  In all the
fits, we used  the {\it tbabs} model and the  ISM abundances (Wilms et
al.\ 2000) to account for the effects of interstellar absorption.  The
XMM-Newton  Calibration  team  reported  that the  EPIC-MOS  detectors
measure  5-8\% higher fluxes  than the  EPIC-pn consistently  over the
whole            energy            band\footnote{see,            e.g.,
  http://xmm.vilspa.esa.es/docs/documents/CAL-TN-0018.pdf.}.  However,
the  exact amount  of this  excess flux  varies over  different energy
ranges.   Therefore, in  all  our fits  we  allowed the  normalization
parameters of the models to be  free between the detectors and added a
1\% systematic  uncertainty to take  into account the  possible energy
dependence  of  the uncertainties  in  instrumental calibrations.   We
calculated the unabsorbed  flux using the {\it cflux}  model in XSPEC.
Because of the above-mentioned calibration uncertainty, we only report
the flux  and/or emitting  area values as  measured with  EPIC-pn.  We
assumed  a  gravitational  redshift  for  the neutron  star  as  0.25,
corresponding to a  neutron star mass of 1.4  M$_{\sun}$ and radius of
11.5  km. We  fit all  spectra in  the 0.5  to 6.5  keV  energy range.
Unless  otherwise noted,  all the  uncertainties quoted  are  for 68\%
confidence interval.

\begin{figure*}
\centering
\includegraphics[scale=0.38, angle=270]{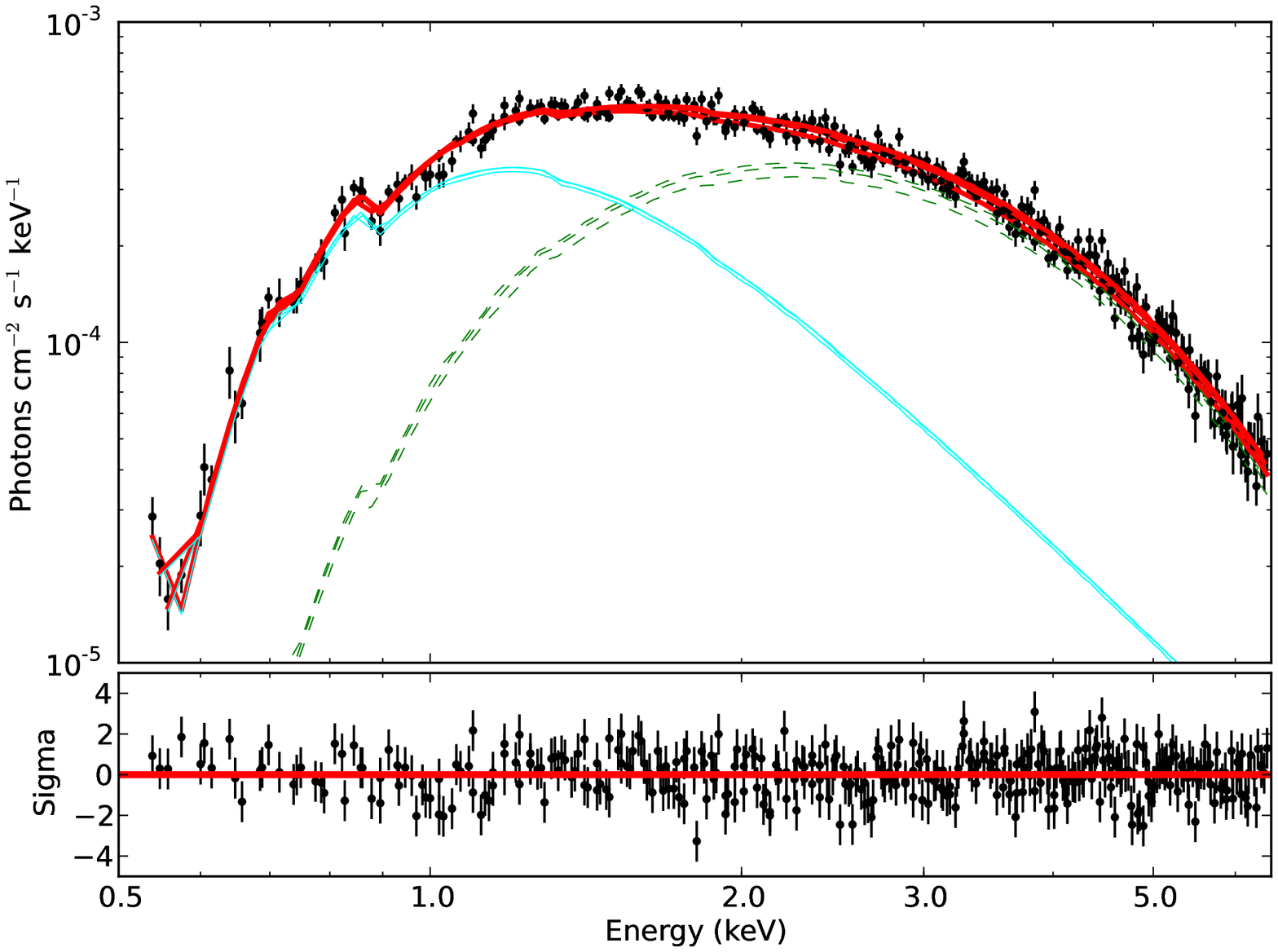}
\includegraphics[scale=0.38, angle=270]{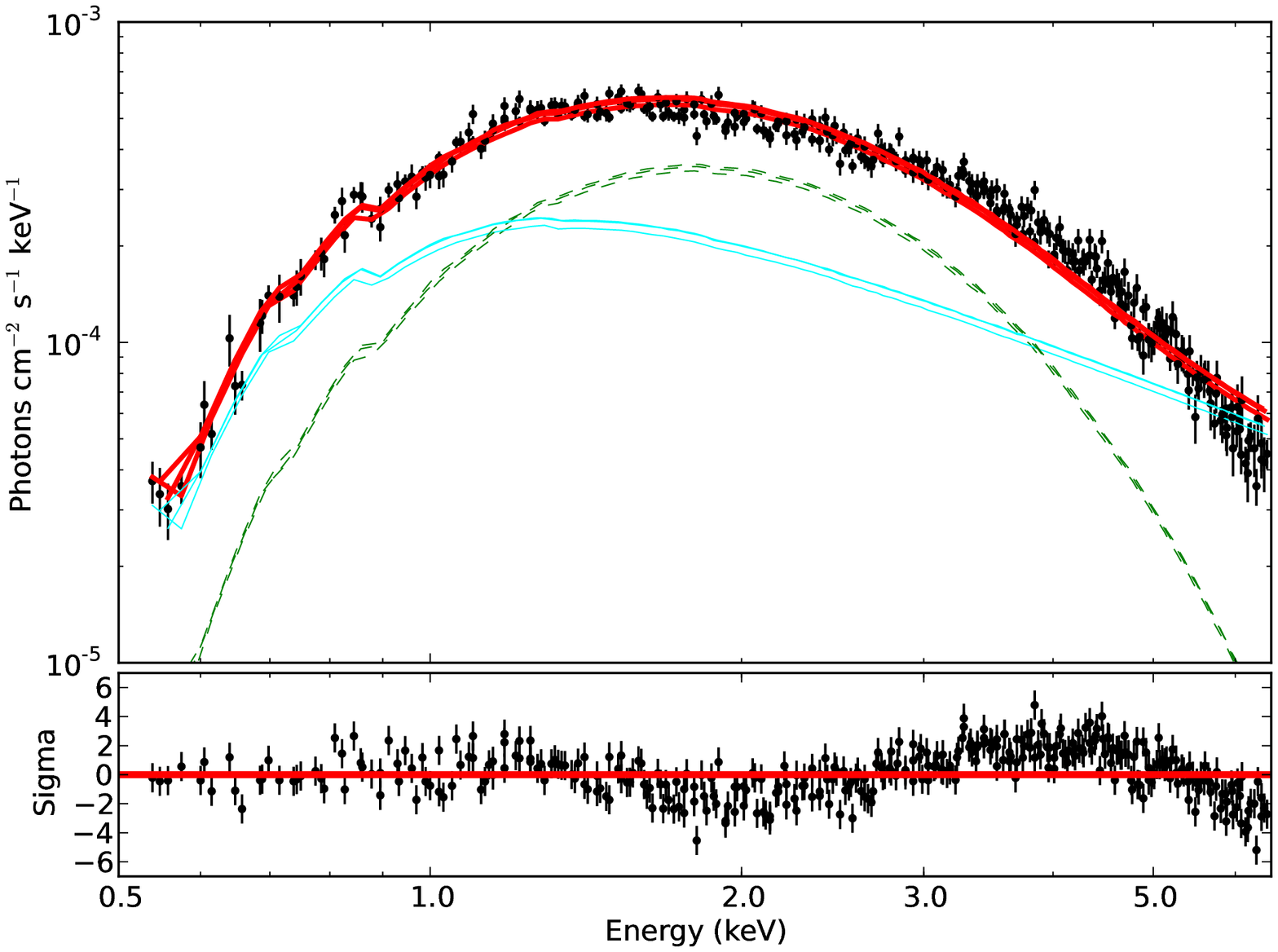}
\caption{Best fit fluxed blackbody  + power-law (upper left panel) and
  neutron star atmosphere (NSA) + power-law (upper right panel) fit to
  the  X-ray spectrum  of  \sgr\ obtained  with  the XMM-Newton.   The
  magnetic  field strength  in the  NSA model  is set  to $10^{12}$~G.
  Blackbody +  power-law provides a  moderate fit with $\chi^2  / {\rm
    dof} = 1.12$ for 346 degrees of freedom, while the NSA + power-law
  is a poor  fit with $\chi^2 /  {\rm dof} = 2.78$ for  346 degrees of
  freedom.  In both cases, the thermal component dominates (shown with
  dashed green  lines) over the power-law component  (shown with thick
  cyan lines) at high photon  energies, i.e., in the $>1.5$~keV range,
  which is  difficult to interpret physically.  Lower  panels show the
  residuals from each fit, respectively.}
\label{unfolded}
\end{figure*}

We  first fit the  X-ray spectra  using a  combination of  a blackbody
({\it  bbodyrad} as  defined  in  XSPEC) and  a  power-law model.   We
performed this  analysis only as a  phenomenological classification of
the source spectrum because the  atmospheric structure even for a weak
or ``zero'' magnetic  field has a strong effect  on the spectral shape
of the X-ray  emission of neutron stars and has  been shown to distort
it  away from  a blackbody  (see e.g.   Romani\ 1988;  Pavlov  et al.\
1995). The spectrum and the best-fit model are shown in the left panel
of  Figure~\ref{unfolded}.    Blackbody  plus  power-law   provided  a
moderate  fit  with  a  $\chi^{2}$/dof  of 1.12  for  346  degrees  of
freedom. The best  fit model resulted in a  hydrogen column density of
$(1.10 \pm 0.05) \times  10^{22}$~cm$^{2}$, a blackbody temperature of
$0.93 \pm 0.006$~keV, and a photon index of the power-law component of
$3.18 \pm  0.19$. The temperature  is unusually high for  two reasons.
First, the blackbody component  dominates over the power-law component
at high photon energies, i.e., in the $>1.5$~keV range, as can be seen
in Figure~\ref{unfolded},  which is physically  difficult to interpret
and is  also contrary to  what is seen  in the spectra of  other SGRs.
Second,  this  high  temperature  corresponds to  an  extremely  small
emitting  radius of  0.18~km/2kpc.  Such a  small  emitting radius  is
especially  hard to interpret  given the  observed pulsed  fraction of
20-40\% in the soft X-rays (Esposito et al.\ 2010).

The current constraints  on the period derivative of  \sgr\ point to a
dipolar magnetic  field strength smaller than  $7.5 \times 10^{12}$~G.
We,  therefore,  tried  to fit  the  spectrum  of  \sgr\ with  low  to
intermediate magnetic  field strength neutron-star  atmosphere models.
The NSA model (Pavlov et  al.  1995; Zavlin, Pavlin, \& Shibanov 1996)
provides  a  model for  the  X-ray  spectra  emitted from  a  hydrogen
atmosphere  of  a  neutron  star  at three  different  magnetic  field
strengths:  $B  <  10^{8} -  10^{9}$~G,  $B  =  10^{12}$~G, and  $B  =
10^{13}$~G.  The {\bf fit} parameters of the NSA model, in addition to
the magnetic field strength, are the neutron star mass and radius, the
surface temperature, and the normalization, which is a function of the
source distance $1/D^2_{\rm kpc}$.

We consider  the $B<  10^8$~G case only  for completeness,  because it
would be unfeasible to account  for the X-ray pulsations observed from
\sgr\ if indeed possessed a  negligible magnetic field. For this case,
the NSA model does not provide  a good fit with a $\chi^{2}$/dof equal
to  1.29 for  350 degrees  of freedom  and yields  a  best-fit surface
temperature  of   0.73~keV  ($8.5   \times  10^6$~K).   As   with  the
blackbody-plus-power-law fit, this rather high temperature corresponds
to a  very small best fit  normalization of $3.0  \times 10^{-10}$, or
equivalently an emitting radius of $R = 0.4$~km/2~kpc.

At a  field strength  of $B=10^{12}$~G, the  NSA model cannot  fit the
spectrum of \sgr, yielding a  minimum $\chi^{2}/{\rm dof} = 13.88$ for
350 degrees  of freedom.  The surface temperature  also hits $10^7$~K,
which is the maximum of the  allowed range in the models. Finally, the
obtained best fit normalization of $8.5 \times 10^{-11}$ translates to
an  unphysically small emitting  radius of  $R =  0.21$~km/2~kpc. This
indicates that  the spectrum of  \sgr\ is inconsistent with  a surface
magnetic  field  strength  that  is  comparable to  its  dipole  field
strength inferred from its period derivative.

We performed a final NSA fit  where we set the magnetic field strength
to $B =  10^{13}$~G.  The quality of the fit  improved compared to the
$10^{12}$~G case but is  still not adequate, yielding a $\chi^{2}/{\rm
  dof} = 3.88$  for 350 degrees of freedom.   The best-fit temperature
again  hits  the  upper  limit   of  the  allowed  range  at  $10^7$~K
(0.86~keV), corresponding to a normalization of $1.3 \times 10^{-10}$,
which is an emitting radius of $R=0.3$~km/2~kpc.

In an attempt  to obtain statistically better fits  and to investigate
whether  additional spectral components  would have  an effect  on the
measured  effective  temperature values,  we  also  modeled the  X-ray
spectrum  of \sgr\ with  an absorbed  NSA plus  a power-law  model. We
found  that  the addition  of  the  power-law component  statistically
improved the  fits, yielding $\chi^{2}$/dof values of  1.13, 2.78, and
1.49 for 346 degrees of  freedom, when the magnetic field strength was
set  to 0,  10$^{12}$,  and 10$^{13}$~G,  respectively.  However,  the
addition  of the  power-law component  did not  decrease  the inferred
effective  temperature, resulting  in  values that  still reached  the
upper  temperature  limit  of  the  model.   In  the  right  panel  of
Figure~\ref{unfolded}, we  show the X-ray spectra and  an example case
for the  NSA+PL models, where the  strength of the  magnetic field was
set to $10^{12}$~G.

We finally modeled the spectrum of \sgr\ with magnetar strength fields
in  the few $\times  10^{13}-10^{15}$~G range,  which may  reflect the
higher multipole  field strengths present at the  stellar surface.  To
this  end, we  used the  Surface Thermal  Emission  and Magnetospheric
Scattering model (STEMS, see G\"uver  et al.  2007, 2008). STEMS takes
into account  the effects  of the ultrastrong  magnetic fields  on the
fully ionized  hydrogen atmospheres  of magnetars (\"Ozel  2001, 2003)
and the resonant cyclotron scattering of the surface photons by mildly
relativistic  charges  in   the  magnetosphere  (Lyutikov  \&  Gavriil
2006). For the  calculation of the surface emission,  we follow \"Ozel
(2001,  2003)  and  solve   the  radiative  transfer  equations  in  a
polarization-mode-dependent manner  including the absorption, emission
and scattering processes in the  atmosphere. We also take into account
the  effects   of  vacuum  polarization  resonance   and  include  the
interaction of photons with protons in the plasma, which gives rise to
absorption features at the  proton cyclotron energy (\"Ozel 2003).  In
the  stellar magnetosphere,  we  incorporate a  treatment of  resonant
cyclotron scattering, following  Lyutikov \& Gavriil (2006).  Resonant
cyclotron scattering can take place in a neutron star magnetosphere as
long  as there  is  a sufficient  density  of moderately  relativistic
electrons and the resulting  up-scattering shifts the initial spectrum
to  higher  energies and  smears  out  the  proton cyclotron  features
(Lyutikov \& Gavriil 2006).

In the STEMS  model, we use the emission emerging  from the surface as
input  for  the resonant  cyclotron  scattering  model  to obtain  the
resulting energy  distribution of photons.  In total  STEMS depends on
four  parameters:  the  surface  effective  temperature  ($kT  =  0.1-
0.6$~keV), the strength of the magnetic field at the surface ($B = 0.6
-  50 \times  10^{14}$~G),  the  optical depth  to  scattering in  the
magnetosphere ($\tau = 1.0 - 12.0$), and the velocity of the particles
in the magnetosphere ($\beta = 0.1 - 0.7$).  Note that, even though in
the calculations of  Lyutikov \& Gavriil (2006), a  high optical depth
is motivated  by a  twist angle, this  is not a  necessary assumption.
Such  high charge  densities are  seen in  the magnetospheres  of even
normal pulsars and the only difference between the \sgr ~and a neutron
star with a magnetar-strength dipole  field would be the occurrence of
the  resonant  layers closer  to  the  neutron  star surface.   As  an
example, based on  the current limit on the dipole  field of \sgr, the
resonant layer would be at  4~R$_{NS}$, whereas for a quadrupole field
that is  10$^{14}$~G at  the surface, the  resonant layer would  be at
$\approx$6~R$_{NS}$.    Finally   we   take   into   account   general
relativistic effects on  the propagation of the photons  by assuming a
gravitational redshift.

\begin{figure*}
\centering
\includegraphics[scale=0.35, angle=270]{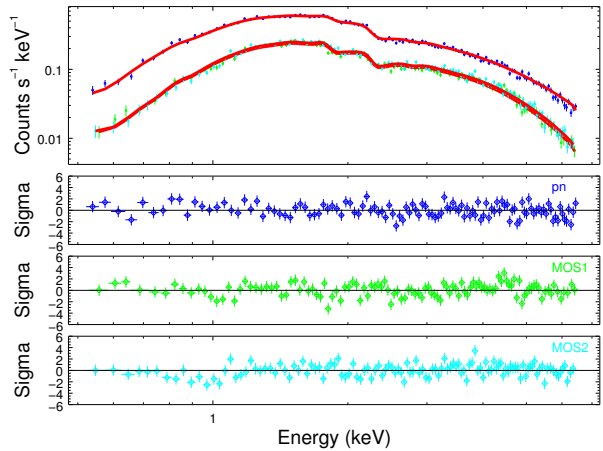}
\caption{XMM-Newton  EPIC-pn  (blue)  and  EPIC-MOS (green  and  cyan)
  spectra of the SGR~0418$+$5729.  Best  fit STEMS model is also shown
  with red thick lines. Residuals from the model for each detector are
  shown in lower panels.}
\label{spectra}
\end{figure*}

Compared to the  blackbody plus a power-law fit,  STEMS model provided
an  equally  good fit  to  the  XMM-Newton  spectrum, resulting  in  a
$\chi^{2}$/dof of 1.18  for 347 degrees of freedom.   We show the best
fit model  curve and the  X-ray spectra in  Figure~\ref{spectra}.  The
parameters  of the  best fit  model were:  the surface  magnetic field
strength  $B = (1.00_{-0.01}^{+0.02})  \times 10^{14}$~G,  the surface
effective     temperature    $kT~=~0.246_{-0.01}^{+0.003}$~keV,    the
scattering    optical   depth   in    the   magnetosphere    $\tau   =
8.94_{-0.18}^{+1.18}$ and  the particle velocity  in the magnetosphere
$\beta  = 0.56\pm0.01$.  The  inferred optical  depth and  the average
electron  velocity are  well within  the assumptions  of  the resonant
cyclotron  scattering model (see.   e.g., Section  3.4 of  Lyutikov \&
Gavriil 2006).  We also found  a hydrogen column density of $(0.74 \pm
0.02) \times  10^{22}$~cm$^{2}$ and the unabsorbed  0.5-6.5 keV source
flux as $(8.53_{-0.06}^{+0.05})\times10^{-12}$~erg~s$^{-1}$~cm$^{-2}$.
The   inferred  emitting  radius   corresponding  to   this  effective
temperature and flux  is $R=2.98$~km/2~kpc.  We note that  in this fit
fractional  emitting  area,  i.e.,  $A_{\rm  hot}/A_{\rm  NS}$,  where
$A_{\rm NS}$ is the entire neutron  star surface area, is in the right
range  to produce  the  pulsed  fraction of  20-40\%,  as observed  by
Esposito et al.\ (2010) in the soft X-rays.

\begin{figure*}
\centering
\includegraphics[scale=0.25]{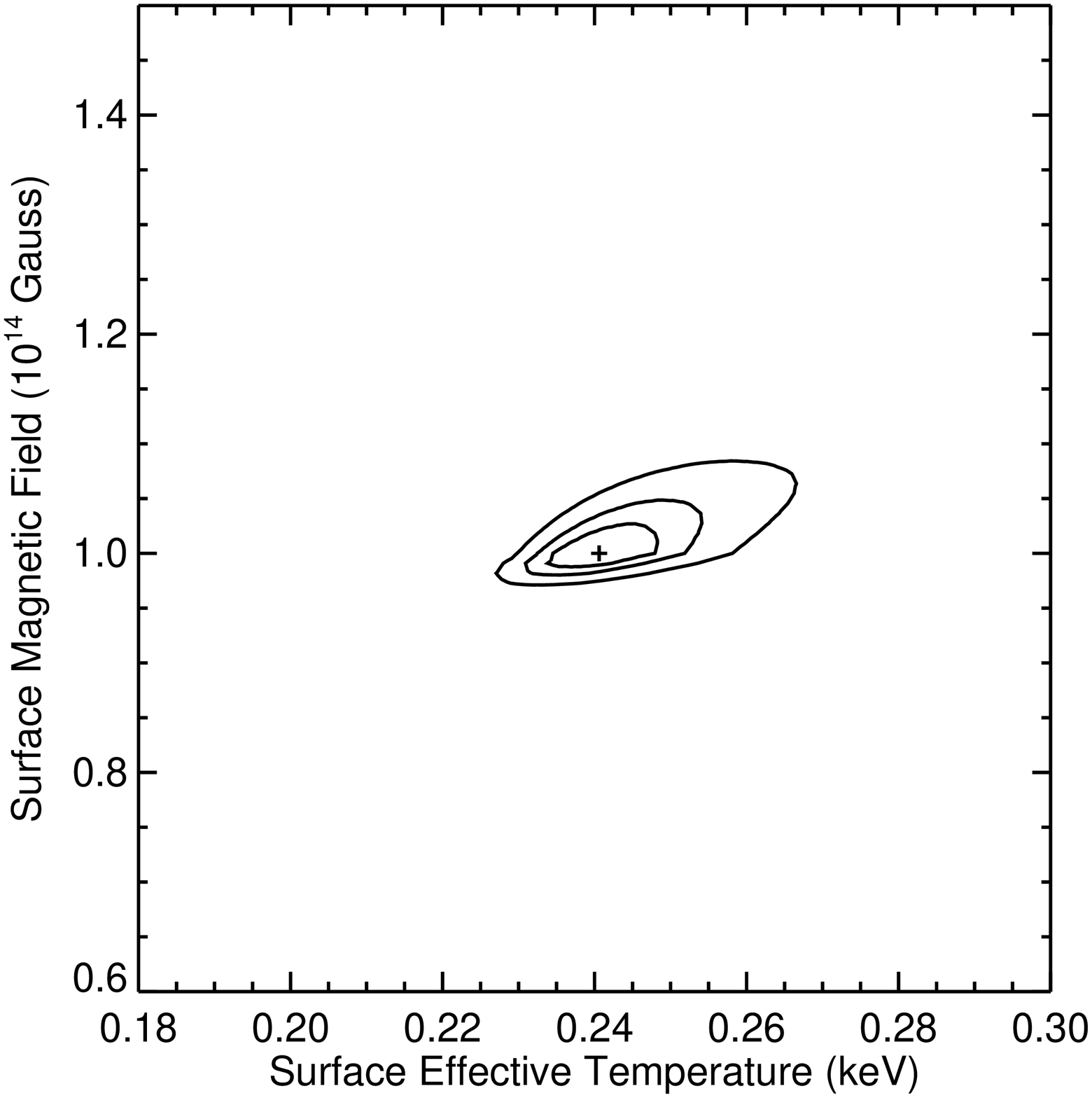}
\includegraphics[scale=0.25]{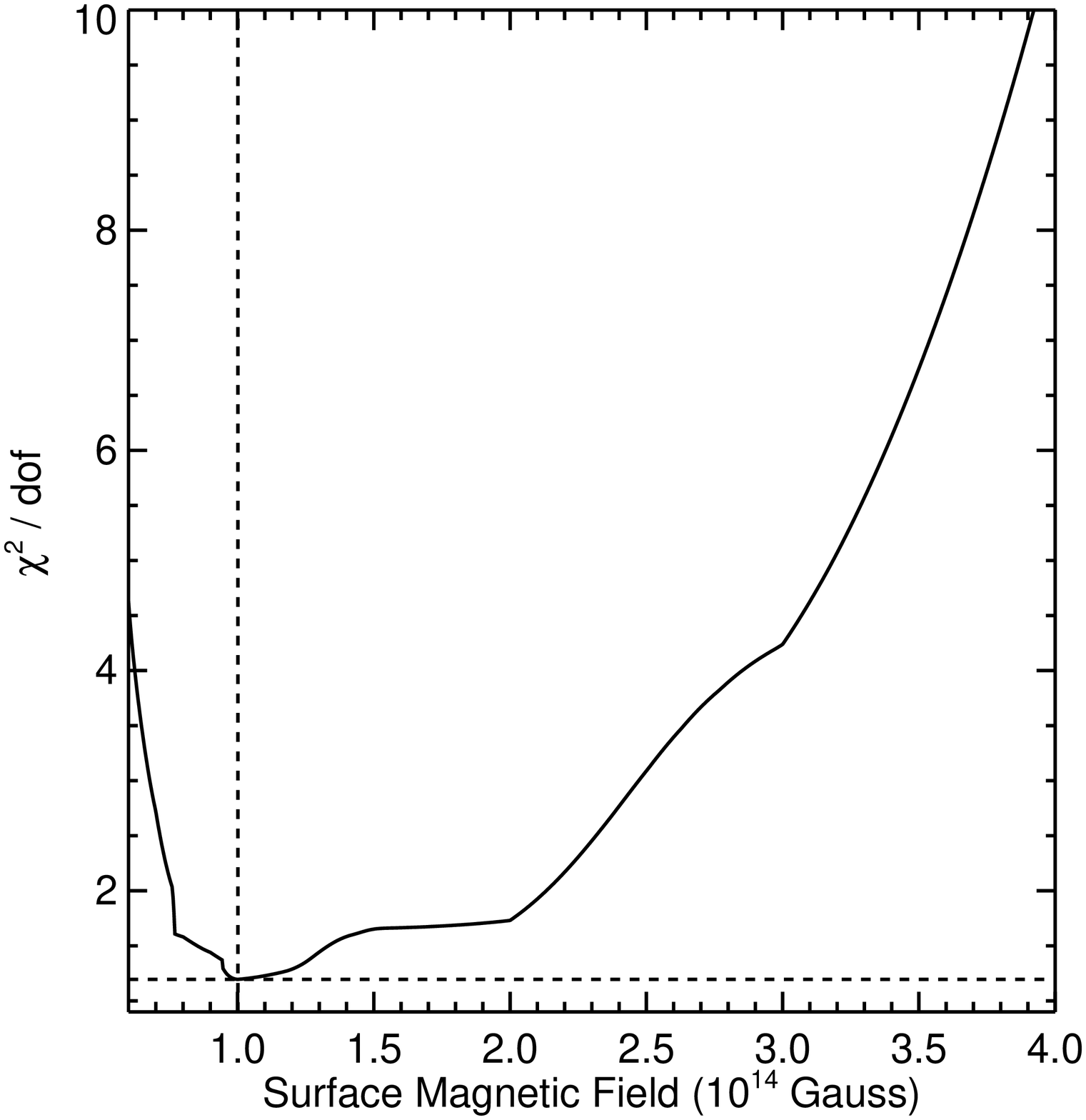}
\caption{{\it Left Panel:} Confidence contours of the best fit surface
  effective temperature and the magnetic field strength at the surface
  of the  neutron star as inferred  from the STEMS  model.  {\it Right
    Panel:} Variation of the $\chi^{2}  / {\rm dof} $ with the surface
  magnetic field  strength. Horizontal and vertical  dashed lines show
  $\chi^{2}$/dof and surface magnetic  field strength for the best fit
  model.}
\label{confidence}
\end{figure*}

We present  in Figure~\ref{confidence} the confidence  contours of the
best fit effective temperature and  the magnetic field strength at the
surface  of the  neutron star.   From the  confidence contours  it can
clearly be  seen that the  STEMS model provides  a lower limit  to the
surface  magnetic field  strength of  the neutron  star of  $\approx 1
\times  10^{14}$~G. Because  our focus  in this  paper is  the surface
magnetic field  strength of \sgr, we further  explored the uncertainty
in this parameter by doing the  following: We fit the data by freezing
the magnetic field at 1000 different points within the $0.6-4.0 \times
10^{14}$~G  range   and  allowing   the  other  parameters   to  vary.
Figure~\ref{confidence} shows the variation of the $\chi^{2}$/dof over
the full  magnetic field range  investigated.  It is evident  from the
distinct  minimum in  the $\chi^2$  in this  figure that  the magnetic
field  strength  at  the  surface  of the  neutron  star  is  uniquely
constrained and is found to be $1.0 \times 10^{14}$~G.

\section{Discussion}

The  recent discovery and  the subsequent  observations of  \sgr\ have
raised  a number of  questions on  our understanding  of the  SGRs and
AXPs,  especially owing to  a lack  of secular  evolution in  its spin
period.  A number of explanations, both within and outside the context
of the magnetar model, have been proposed to interpret the peculiarity
of \sgr.   For example,  the effects of  a stronger  toroidal magnetic
field  below the surface  on the  neutron star  crust was  employed to
create the observed X-ray bursts independent of strength of the dipole
component  (see, e.g., Perna  \& Pons  2011; Rea  et al.\  2010).  The
existence of a  fallback disk has been proposed  to evolve the neutron
star  to  its current  spin  period  with a  dipole  field  as low  as
10$^{12-13}$~G (Alpar, Ertan, \&  {\c C}al{\i}{\c s}kan 2010; see also
Ertan et al.\ 2007).

In this  paper, we analyzed the  X-ray spectrum of  \sgr\ to constrain
its surface  magnetic field strength,  as well as the  temperature and
the magnetospheric parameters  of the neutron star.  We  find that the
empirical blackbody + power-law model provides a moderate fit, but the
emitting  area  inferred from  this  fit  is  unphysically small  ($R=
0.18$~km/2~kpc).  In  addition, contrary to  the physical expectations
and in contrast to other  AXP and SGR spectra, the blackbody component
dominates over  the power-law component  in the higher, as  opposed to
the  lower, energy regime.   We also  found that  realistic atmosphere
models of a neutron star with moderate magnetic field strengths (NSA),
do not  describe the  spectrum adequately, yielding  $\chi^2/{\rm dof}
\geq  3.9$.    In  contrast,  neutron  star   atmosphere  models  with
magnetar-strength fields  (STEMS) produces a  fit with $\chi^{2}$/{\rm
  dof} $=$1.18  for 347  degrees of freedom  degrees of  freedom.  The
best  fit  value  of  the  magnetic  field  strength  is  $1.0  \times
10^{14}$~G.   Thus,  the spectral  analysis  strongly disfavors  X-ray
models where the magnetic field  strength at the surface is assumed to
be 10$^{8-9}$, 10$^{12}$, or 10$^{13}$~G.

\begin{figure*}
\centering
\includegraphics[angle=270,scale=0.5]{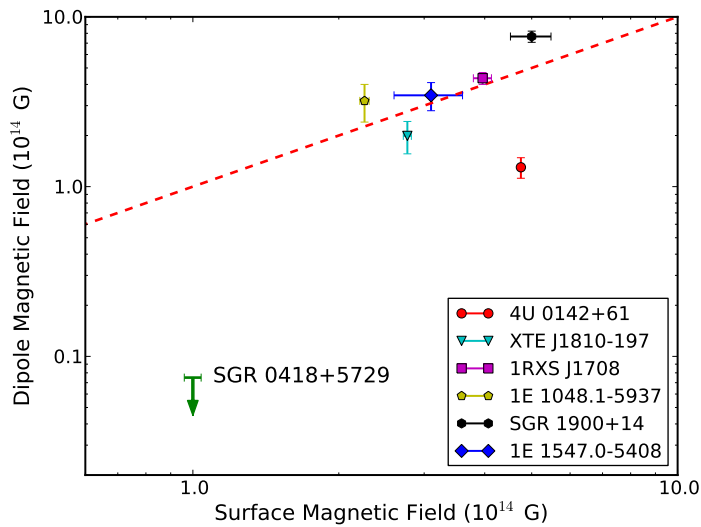}
\caption{Comparison of the magnetic  field strengths as inferred using
  the  STEMS model  to the  dipole  fields deduced  from the  spindown
  properties for seven magnetar  candidates.  The error bars in dipole
  field strengths  represent the range of measured  spindown rates for
  each  source, while  the error  bars in  the  spectroscopic magnetic
  field      strength       represent      2-$\sigma$      statistical
  uncertainties.    Dashed    lines    show   the    relation    where
  $B_{surf}=B_{dip}$.}
\label{comp}
\end{figure*}

X-ray  observations   of  a  number  of  other   magnetars  have  been
successfully modeled with neutron  star atmospheres with high magnetic
field  strengths, including  XTE~J1810$-$197 (G\"uver  et  al.\ 2007),
4U~0142+61     (G\"uver    et     al.\     2008),    1E~1048.1$-$5937,
1RXS~J170849.0$-$400910 (\"Ozel et al.\ 2008), 1E~1547.0$-$5408 (Ng et
al.\ 2011) and SGR~1900+14  (G\"o\u{g}\"u\c{s} et al.\ 2011), and were
used to  obtain the  surface parameters of  the neutron stars  such as
their  magnetic  field  strengths  and  effective  temperatures.   The
spectroscopically  inferred magnetic field  strength values  for these
sources, in  units of  10$^{14}$ G, are  2.77$\pm$0.05, 4.75$\pm$0.03,
2.26$\pm$0.05,    3.96$\pm0.17$,   3.1$\pm$0.5,    and   5.0$\pm$0.48,
respectively.     In    Figure    \ref{comp},   we    compare    these
spectroscopically  determined  field  strengths  with  the  previously
reported  inferred dipole  magnetic field  for each  source,  which we
obtained           from          the           McGill          SGR/AXP
Catalog\footnote{http://www.physics.mcgill.ca/~pulsar/magnetar/main.html},
Ng  et al.\  (2011), and  \"Ozel  et al.\  (2008).  For  all of  these
sources, the spectrally inferred  surface magnetic field strength {\bf
  is}  in  good  agreement   with  dipole  magnetic  field  estimates,
differing at  most by a factor  of four.  \sgr\ is  the first magnetar
candidate   for  which   the  surface   magnetic  field   strength  is
significantly  larger than  the  limit on  the  dipole magnetic  field
($\ge$~15  times larger).   It is  also interesting  that  the surface
magnetic field strength  we report here for \sgr\  is the lowest among
other STEMS measurements for other AXPs and SGRs so far.

Even though the dipole  field strengths inferred from spindown involve
a  number  of  simplifying  assumptions, relaxing  them  in  realistic
simulations  leads to  a  factor  of two  difference  in the  inferred
magnetic field strengths (Contopoulos \& Spitkovsky 2006).  Therefore,
the discrepancy between the  field strength inferred from the spindown
of \sgr\  and the field strength  inferred from both  its spectrum and
its bursts is too large to be accounted for in this way.  Instead, the
difference points to a complex magnetic field geometry, i.e, it can be
attributed to the presence of higher order multipole components at the
surface of  the neutron star,  which can shape the  characteristics of
the  X-ray  emission but  do  not  contribute  to the  spindown.   The
multipole fields  are likely to play  a role in  determining the pulse
shapes   observed  in   AXPs  and   SGRs  (\"Ozel   2002).    In  this
interpretation, the higher order  components of the magnetic field can
also cause  the fracturing of the  neutron star crust,  leading to the
observed X-ray  bursts. Further comparisons of the  surface and dipole
magnetic  field strengths  of  AXPs  and SGRs,  as  well as  numerical
simulations of the magnetic field evolution in young neutron stars may
help further constrain the  magnetic field strengths and geometries of
X-ray bright neutron stars.

\section*{Acknowledgments}  

We thank the anonymous referee  for his/her comments that improved the
clarity  of  the manuscript.   This  work  makes  use of  observations
obtained with XMM-Newton, an  ESA science mission with instruments and
contributions  directly funded  by ESA  Member States  and  NASA.  F.\
\"O. and T.\ G.\ acknowledge support from NSF grant AST-07-08640.

\label{lastpage}

\end{document}